\pdfoutput=1 
\newcommand{\arXiv}[1]{\href{http://www.arxiv.org/abs/#1}{\tt arXiv:#1}}
\documentclass{JINST}
\let\ifpdf\relax
\title{Cosmic Muon Flux Measurements at the\\Kimballton Underground Research Facility}
\author{L. N. Kalousis$^{a,}$\thanks{Corresponding author}, E. Guarnaccia$^a$, J. M. Link$^a$, C. Mariani$^a$ and R. Pelkey$^a$ \\
\llap{$^a$}Center for Neutrino Physics, Virginia Polytechnic Institute and State University, \\
Blacksburg, Virginia 24061, USA \\
E-mail: \email{kalousis@vt.edu}  }
\abstract{In this article, the results from a series of muon flux measurements conducted at the Kimballton Underground Research Facility (KURF), Virginia, United States, are presented.   
The detector employed for these investigations, is made of plastic scintillator bars readout by wavelength shifting fibers and multianode photomultiplier tubes.  
Data was taken at several locations inside KURF, spanning rock overburden values from $\sim$ 200 to 1450 m.w.e. 
From the extracted muon rates an empirical formula was devised, that estimates the muon flux inside the mine as a function of the overburden.   
The results are in good agreement with muon flux calculations based on analytical models and MUSIC. }
\keywords{Cosmic muon flux; Underground sites; Plastic scintillator bars; Wavelength shifting fibers; Multianode photomultiplier tubes; Kimballton Underground Research Facility; Muon flux dependence on the overburden}
\usepackage{amsmath}
\usepackage{amssymb}
\begin{document}
\section{Introduction}\label{sec:intro}
\par High energy muons, created from the collisions of primary cosmic ray particles in the atmosphere, are a serious source of background for many detectors probing the properties of neutrinos, the existence of dark matter and double $\beta$-decay~\cite{joe}.  Besides neutrinos, muons are the only components of cosmic rays that survive at underground sites with significant rock overburden. A good knowledge of the muon flux is necessary for all the current and planned experiments that require low background. Muons spallation can create long lived isotopes while passing through the detector material and they can generate hadronic showers, including fast neutrons in the rocks surrounding the detector. 

\par The Kimballton Underground Research Facility (KURF) is one of the major, deep underground laboratories in the North America. KURF is located at the Kimballton mine in Giles County, Virginia. It is an active limestone mine currently operated by Lhoist of North America~\cite{loist}. The main features of KURF include: the large space available, drive-in access even at its lowest levels, close proximity to Virginia Tech (26~miles from the main campus), and low levels of natural radioactivity ($\sim$15~Bq/Kg for $^{40}$K, $\sim$1.5 Bq/Kg for $^{226}$Ra, $\sim$0.7~Hz for $^{226}$Th and $<$15 Bq/m$^3$ for $^{222}$Rn)~\cite{lens1} due to the rock composition which is mostly limestone and dolomite.  Examples of recent and current experiments at KURF include: Low-Energy Neutrino Spectroscopy (LENS)~\cite{lens,lens1}, Majorana Low-Background Broad Energy Germanium detectors~\cite{malbek}, high-purity germanium (HPGe) detectors to study the double beta decay modes of $^{100}$Mo and $^{150}$Nd to excited states~\cite{tunl1,tunl2}, measurements of the contamination of $^{39}$Ar in pure Argon for future dark matter experiments~\cite{ctf,dark}, and a gamma-counting facility all located at the 1,450~meters water-equivalent (m.w.e.) level.  
KURF also hosts the WATCHBOY~\cite{adam} detector, a prototype for a water-based reactor monitoring program, at a depth of 300~m.w.e.

\par In this article we report on a series of muon flux measurements performed at KURF on various locations spanning depths between 200 and 1,450~m.w.e. A measurement of muon rate at the 1,450~m.w.e. level has been previously published~\cite{dark} and efforts have been pursued to determine the neutron flux at that location~\cite{fn}. The analysis described in this paper decreases substantially the uncertainties of the muon flux measurement. In addition, an empirical formula to estimate the muon flux as function of depth at KURF is derived. The paper is organized as follows: Section~\ref{sec:det} gives a description of the muon detector; Sec.~\ref{sec:flux} gives the results of this flux measurements; and Sec.~\ref{sec:pred} presents the empirical formula used to estimate the muon flux as a function of the rock overburden at KURF.
\section{Muon Detector Description}\label{sec:det} 
\par The detector used for the flux measurements described in this article, was build as part of the Double Chooz effort at Columbia University as a prototype of the Outer Veto system, which is a large muon array deployed above the Double Chooz reactor antineutrino detector~\cite{doublechooz}.
\begin{figure}[t!]
\centering
\includegraphics[width=11.0cm, height=7.6cm]{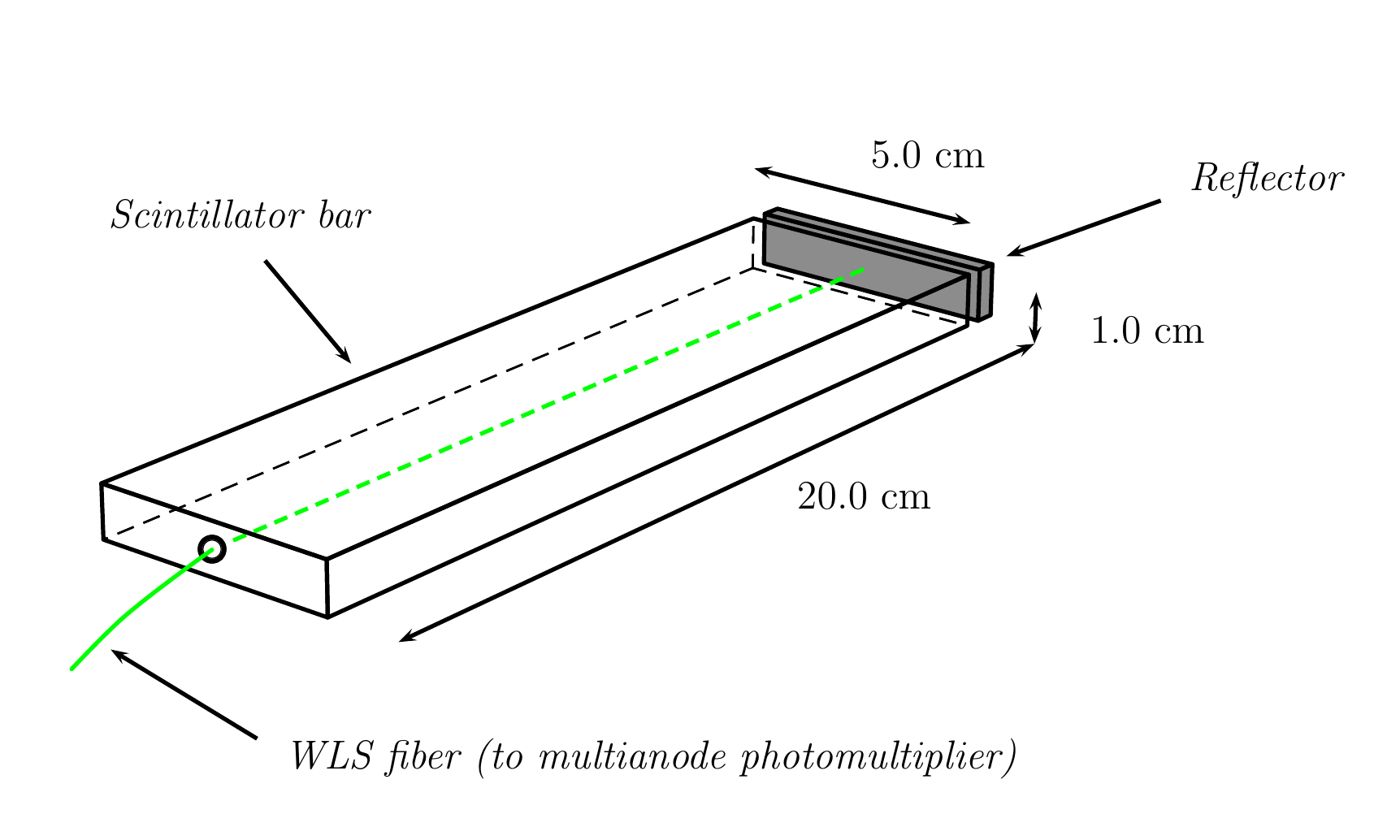}
\caption{Schematic of the plastic scintillator bars.}
\label{fig:bars}
\end{figure}
\par
\begin{figure}[!t]
\centering
\begin{tabular}{c  c  c }
\includegraphics[width=5.2cm, height=3.975cm]{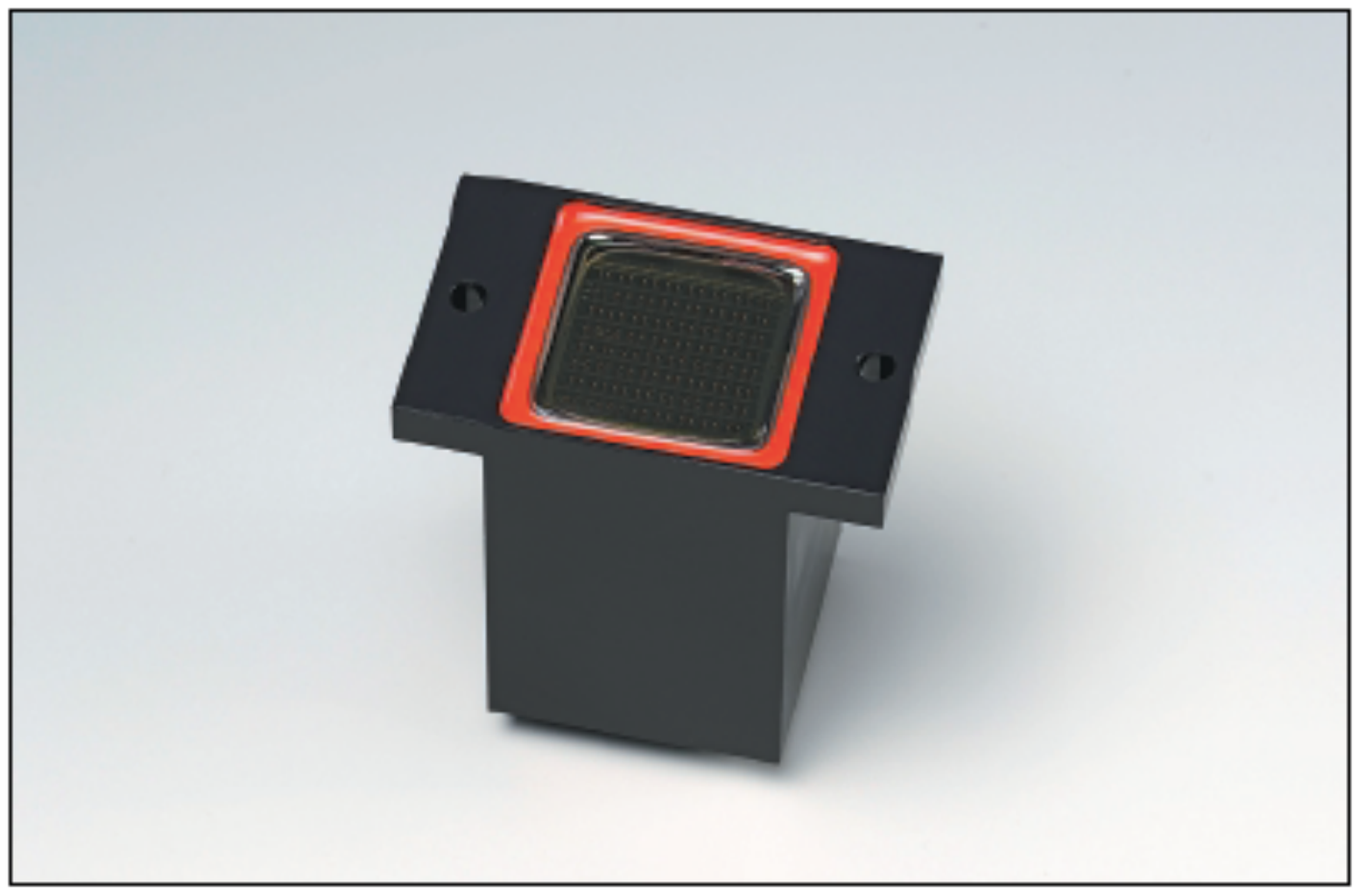}   & \vspace{0.5cm} & \includegraphics[width=5.2cm, height=4.0cm]{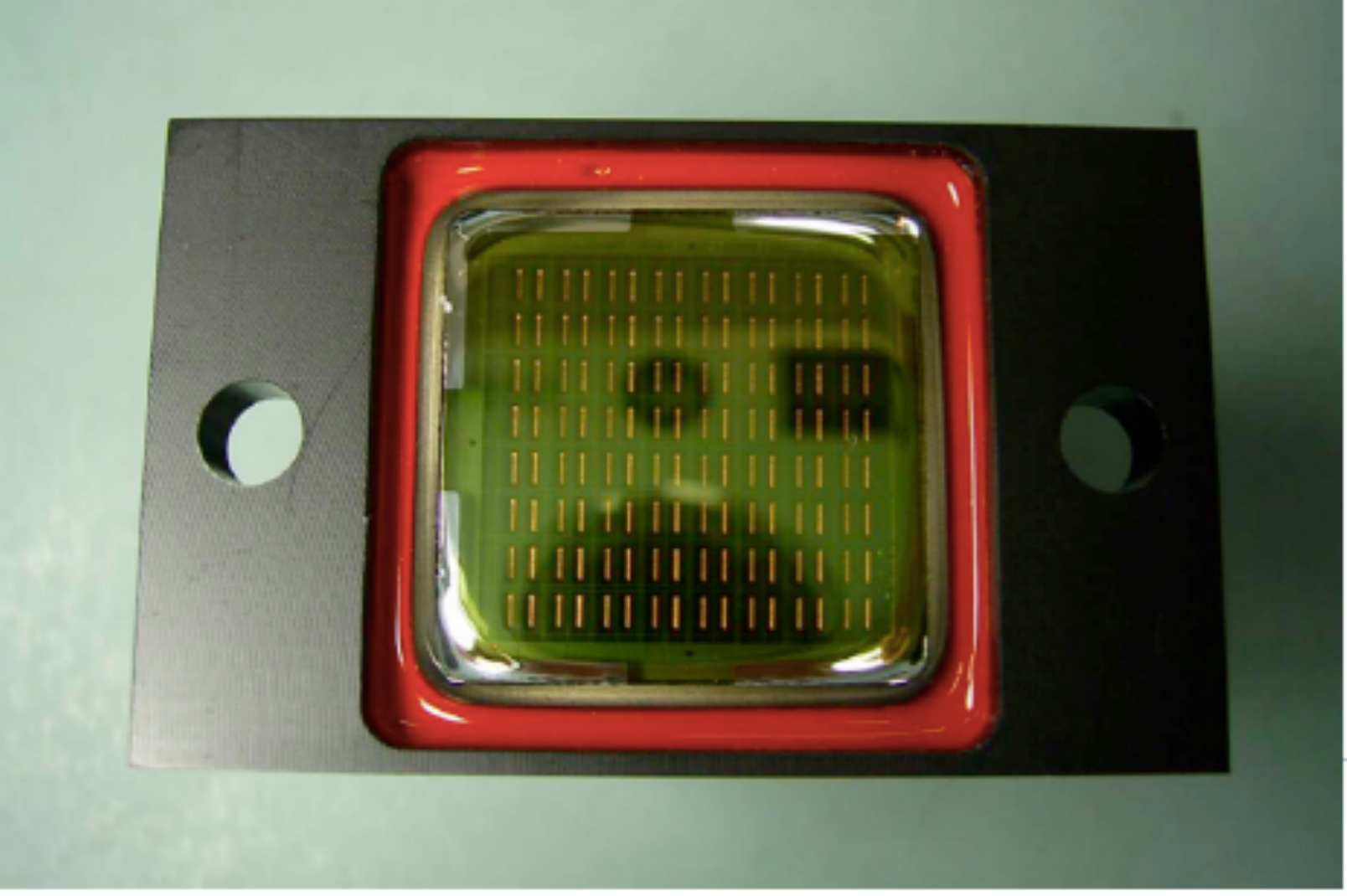} 
\end{tabular}
\caption{The Hamamatsu multianode photomultiplier tube model H8804. }
\label{fig:mapmt}
\end{figure}

 The muon detector is made of plastic scintillator bars of 20~cm in length. The scintillator bars were produced at Fermi National Accelerator Laboratory (FNAL) \cite{fnal, fnal1}. 
Each scintillator bar is made of polystyrene (C$_8$H$_8$), 1.0\% of PPO and 0.03\% of POPOP and it is 20$\times$5$\times$1~cm$^3$. 
Each bar is covered by a 0.25~mm thick reflecting coating made of TiO$_2$ to enhance the reflection of light inside the bars and to ensure optical separation between the bars in the same module. The bars were extruded with a 1.8~mm hole in their center. 
A 1.5~mm diameter wavelength shifting (WLS) fiber (Kuraray Y-11(175 ppm) multi-clad non-S type) is inserted in the hole of each bar to guide the scintillation light to multi-anode photomultiplier tubes (MAPMTs). 
The absorption peak for the fibers is at 430~nm (matching the emission peak of the scintillator), and the emission peak is at 476~nm. 
A schematic of one scintillator bar is shown in Fig.~\ref{fig:bars}. The scintillation light is detected by Hamamatsu H8804 MAPMTs as in Fig.~\ref{fig:mapmt}. 
\begin{figure}[t!]
\centering
\begin{tabular}{c  c  c }
\includegraphics[width=6.2cm, height=6cm]{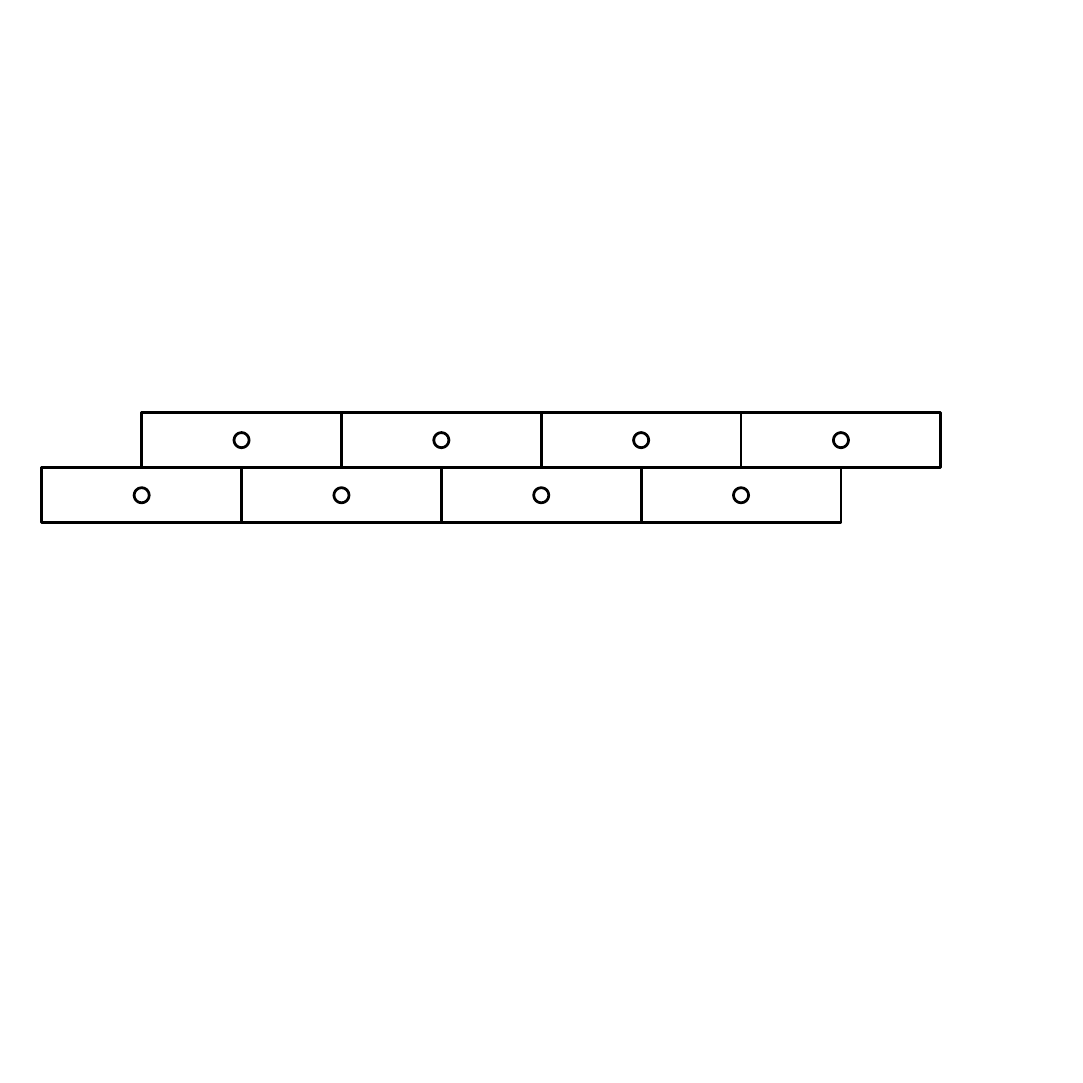} & \vspace*{0.1cm} & \includegraphics[width=8cm, height=6cm]{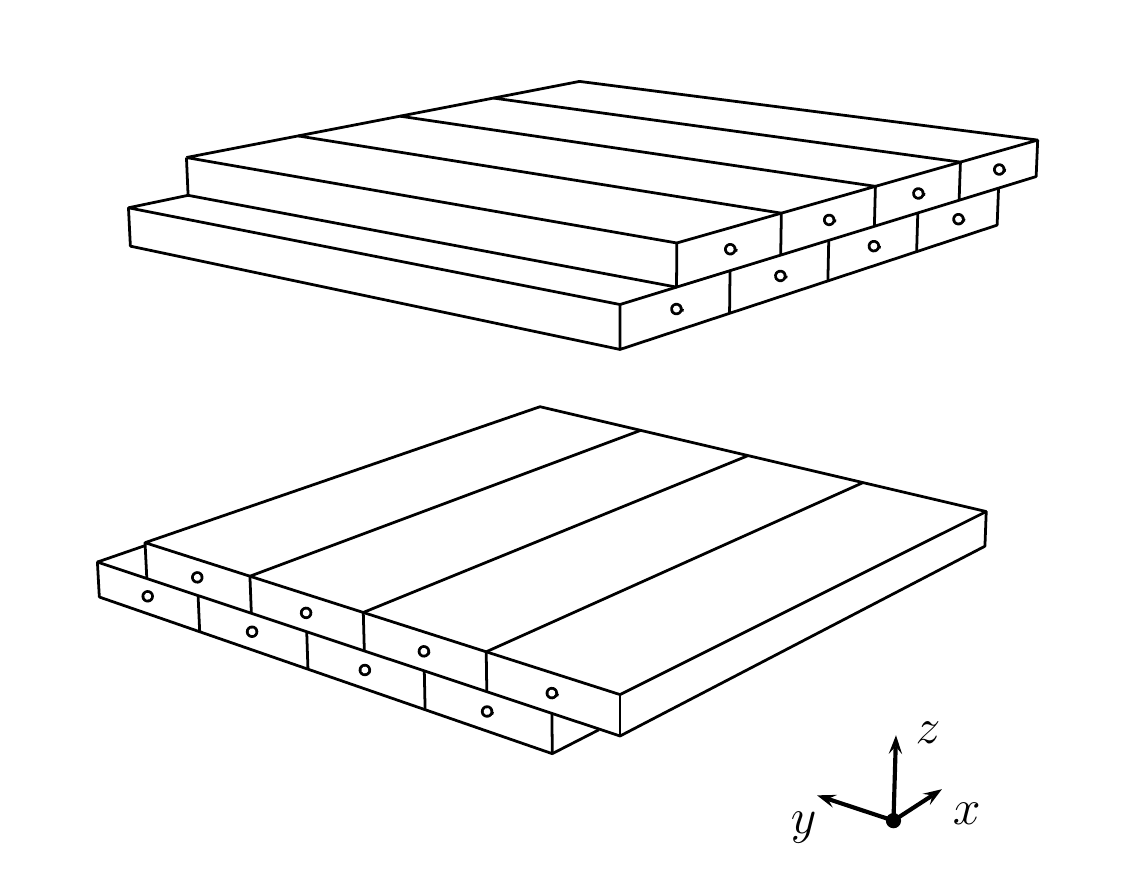}
\end{tabular}
\includegraphics[width=10.5cm, height=4.6cm]{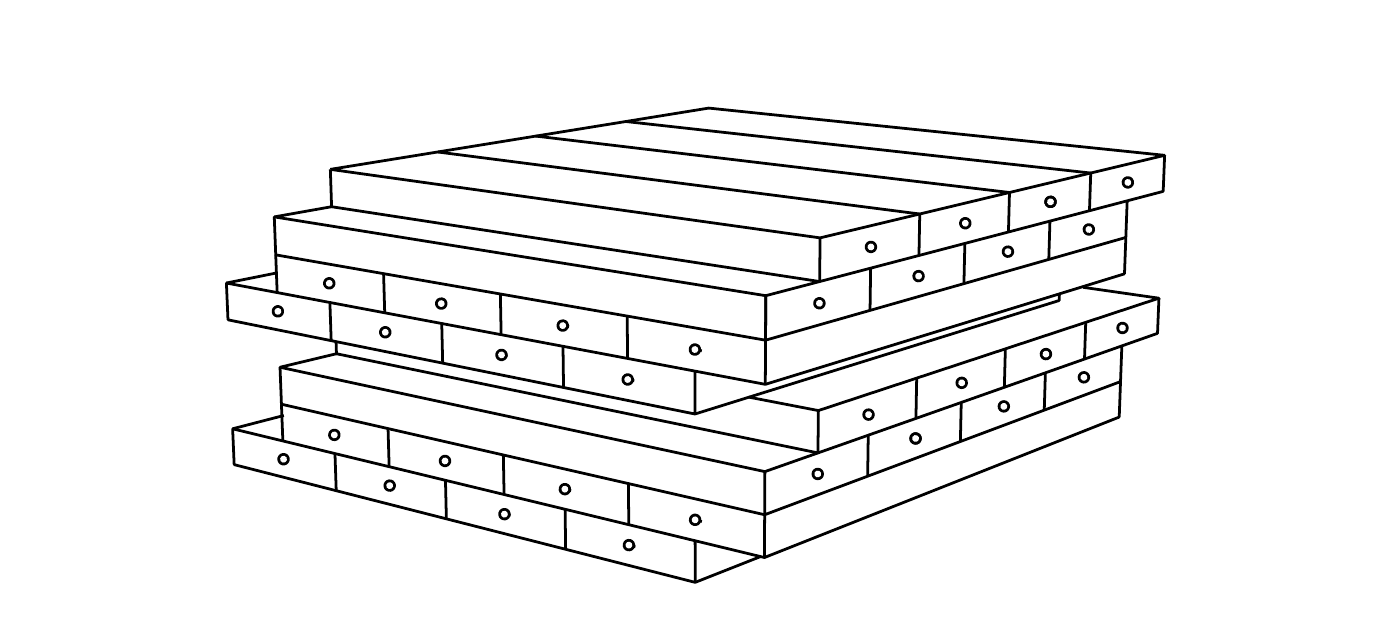}
\caption{Cross-sectional view of the bilayers used (top left) and schematic of the $x$ - $y$ modules (top right). A schematic of the final detector is also shown (bottom). }
\label{fig:detector_assembly}
\end{figure}
%
\par The readout system consists of a front-end electronics board (FEB) attached to each MAPMT and a back-end USB module directly connected to a PC. 
The muon detector is made of 4 independent modules readout by two MAPMTs. Two vertical adjacent modules are readout by the same MAPMT. Each module is oriented perpendicular to the others and they are made of 8 scintillator bars. The 8 scintillator bars are assembled in 2 separate layers, a top and a bottom layer, made of 4 bars each. The top and bottom layers are displaced by 2.5~cm (half of a strip) to replicate the design of the Outer Veto modules of the Double Chooz experiment. On the top and bottom of each layer a 1~mm thick aluminum sheet is glued to the bars to improve the layer's rigidity. A schematic of each module and how the 4 modules are assembled together is presented in Fig.~\ref{fig:detector_assembly}. 
This particular assembly was chosen to maximize muon detection efficiency and to minimize background from random coincidences in the detector, i.e.,  events with hits that are not produced by through-going muons (dark current, natural  radioactivity). 
It is known, form independent studies, that the probability of those events to create eight overlapping hits in the detector is negligible.  
Each scintillator bar has a 99.5\% efficiency and the total efficiency of a module is 97.5\% when two geometrically overlapping hits are required between the top and bottom layers. The full detector efficiency depends strongly on the event selection requirements and it will be discussed in more detail in Sec.~\ref{sec:event_selection}.
The orientation of the 4 module assembly significantly improves the track reconstruction capabilities of the detector\footnote{Note though that no track reconstruction analysis is included in this publication.}. 
\par The four modules are placed inside a wooden box with approximate dimensions of 100~cm $\times$ 70~cm $\times$ 20~cm. The interior of the box has been dyed black and its light tightness has been secured using plastic covers and dark silicon near its seams. Most of the electronics of the detector are housed inside the box but the power supply units are placed outside to avoid problems with heat dissipation.  The detector is portable and can easily be deployed in an experimental hall or inside a mine.
%
%
The detector employs two MAPMTs. 
Each MAPMT has sixty-four pixels and is connected to 2 modules. 
Each module consists of 8 fibers that are coupled to 8 channels in the MAPMT. 
The total number of channels used for reading out the detector is 32. 
These modules will be referred as the $x_1$, $y_1$, $x_2$ and $y_2$ bilayers throughout the rest of the paper. The fibers are attached to the photocathodes using a fiber holder which guarantees a very precise alignment between the fibers and the MAPMT pixel centers. The MAPMT pixels have a surface area of 2$\times$2~mm$^2$ while the fibers' diameter is 1.5~mm. The fibers are glued together into the fiber holder which is coupled to the MAPMT using screws, and they are arranged in such a way that there is at least a 2~mm distance between every fiber pair. That means that there are no neighboring pixels used around each pixel of the MAPMT that is connected to a WLS fiber. This choice minimizes the optical cross-talk between pixels. 
The precision of the alignment of the fibers and MAPMT's pixels was estimated to be of the order of few mils.

\par The MAPMTs are readout by a custom-made board developed by Columbia University and Nevis Laboratories. The core of the readout boards is the MAROC2 chip that discriminates the sixty-four outputs from the MAPMT \cite{maroc2}. It incorporates a slow-shaper with a track-and-hold for each input and a multiplexed output for an external ADC. The outputs from the MAROC2 are sent to an FPGA that can be programmed to read the module only if a certain hit-pattern is satisfied. The hit-pattern determines which of the ADC outputs are written along with the PMT channel number. The digital charge output is given by a 12~bit ADC Wilkinson. Each MAPMT channel gain can be corrected individually using a 6-bit parameter to take into account the gain variation (gain constants). The gain constants are sixty-four numbers corresponding to the total number of MAPMT pixels and they can be set through the data acquisition (DAQ) system. 
Additionally, another level of calibration is applied to every run sequence during the analysis. This consists in a set of correction factors\footnote{%
Also referred as scaling or rescaling factors since they \emph{rescale} the ADC distributions. 
} 
that they are applied to the ADC values in a two-step algorithm that is described in Section~\ref{sec:event_selection}.

\par The PMT boards are daisy chained using ethernet Cat6 cables and the chain is connected to a readout USB board. The USB is then connected to a standard laptop and the data are recorded using a DAQ software. A Visual Basic/C++ software has been developed to interface the USB readout board with a PC. This software uses the library available on Visual Basic or C++ for the FX-USB chip set.

\par The detector was first deployed in the physics department at Virginia Tech, which is at the elevation of 633~m.  A run sequence of about 15 hours was taken, measuring a flux of 170$\pm$26~Hz m$^{-2}$. 
This measurement is in very good agreement with the prediction of muon flux at this elevation.  The data from this run was also used to determine the rescaling constants, and channel and detector module efficiencies. The detector was then moved to KURF for a series of runs at different depths. Table~\ref{table:data} summarizes the data taking time and the overburden for each of the detector positions.

\begin{table}[t!]
\centering
\begin{tabular}{| c | c |}
\hline
\hspace*{1cm} Rock overburden (m.w.e.)  \hspace*{1cm} & \hspace*{1cm} Run length (hours) \hspace*{1cm} \\ \hline
   0      &  16.6 \\
   292 &   18.7 \\
   307 & 27.0   \\
   328 &  12.9 \\
   340 &  14.2 \\
   348 &  17.5  \\
   358 &  20.9 \\
   362 &  19.5 \\
   377 &  15.0 \\
 1,450 &  582.7 \\
\hline
\end{tabular}
\caption{The data sample used in this analysis.}
\label{table:data}
\end{table}

\par The run length at each position was optimized to achieve a statistical error on the order of 10\% or less. The overburden was determined using a mine survey that was performed in Fall 2012. The experimental errors in the survey combined with the uncertainties in to rock densities at the various locations translates to expected uncertainties of about 5\% on the estimation of the depth in m.w.e. for each location. The sites between 292 and 377~m.w.e. are located on Level 2 of KURF while the deeper location of 1,450~m.w.e. is where the Virginia Tech laboratory is located.

\section{Muon Flux Analysis}\label{sec:flux}
The muon analysis technique proceeds in three separate steps. First, a preliminary sample of muons is selected according to the channel-wise hit information. These events are used to re-calibrate the detector and to compute the final muon rate. The muon rate calculation includes an efficiency correction due to the selection used. Two geometrically overlapping hits in the top and bottom layer of each modules were required. A total of 8 geometrically overlapping hits are required.  This selection efficiency 
was computed using only hits at the high threshold of 4 P.E. as it will be explained in the following sections. The analysis is performed directly on the data format that comes out of the Event Builder using an evolved C++/ROOT based software \cite{root}. 

\subsection{Event selection }\label{sec:event_selection} 
\par The muon event selection requires not only geometrically overlapping strips but also uses the pulse amplitudes for all the channels (ADC). The ADC for each of the channels is computed by taking into account the noise of each channel (baseline) that includes both the electronic and MAPMT noises. The ADC value for each channel is converted into P.E. using a conversion factor of $\sim$17 ADC/P.E. that is constant for all the MAPMT channels and it has a variation of about 10\% between the two MAPMTs used in this setup. These conversion factors were predetermined by illuminating all MAPMT channels with a controlled light source (led laser) and analyzing the recorded spectra assuming Poisson statistics. The ADC to P.E. conversion factor was also cross checked using the output of channels that are not directly instrumented but that can receive light as cross-talk from neighboring instrumented channels. The amount of light in this case is 1 P.E. and that can be used for calibration purposes. A vertical muon passing through the middle of two geometrically overlapping strips release 5 P.E.. This value was obtained using high efficiency muon paddles positioned above and below two geometrically overlapping strips of the detector and averaging  the pulse amplitude, ADC, for those hits in overlapping bars over the full length of the scintillator strips.
The ADC mean value was translated to the number of P.E. using the conversion factor explained in Section~\ref{sec:event_selection}. 
Each channel is further corrected by the calibration constants. To be recorded on disk a hit should be at least above $1/3$ of a P.E. This is a fixed hardware threshold. 
There is no geometrical requirement at the hardware level. Once the data is on disk then both a software threshold and a geometrical requirement are applied. 
Five software P.E. thresholds were used in this analysis: $1/3$, 1, 2, 3 and 4 P.E.
%
\par
\begin{figure}[!t]
\centering
\includegraphics[width=8.4cm, height=6.25cm]{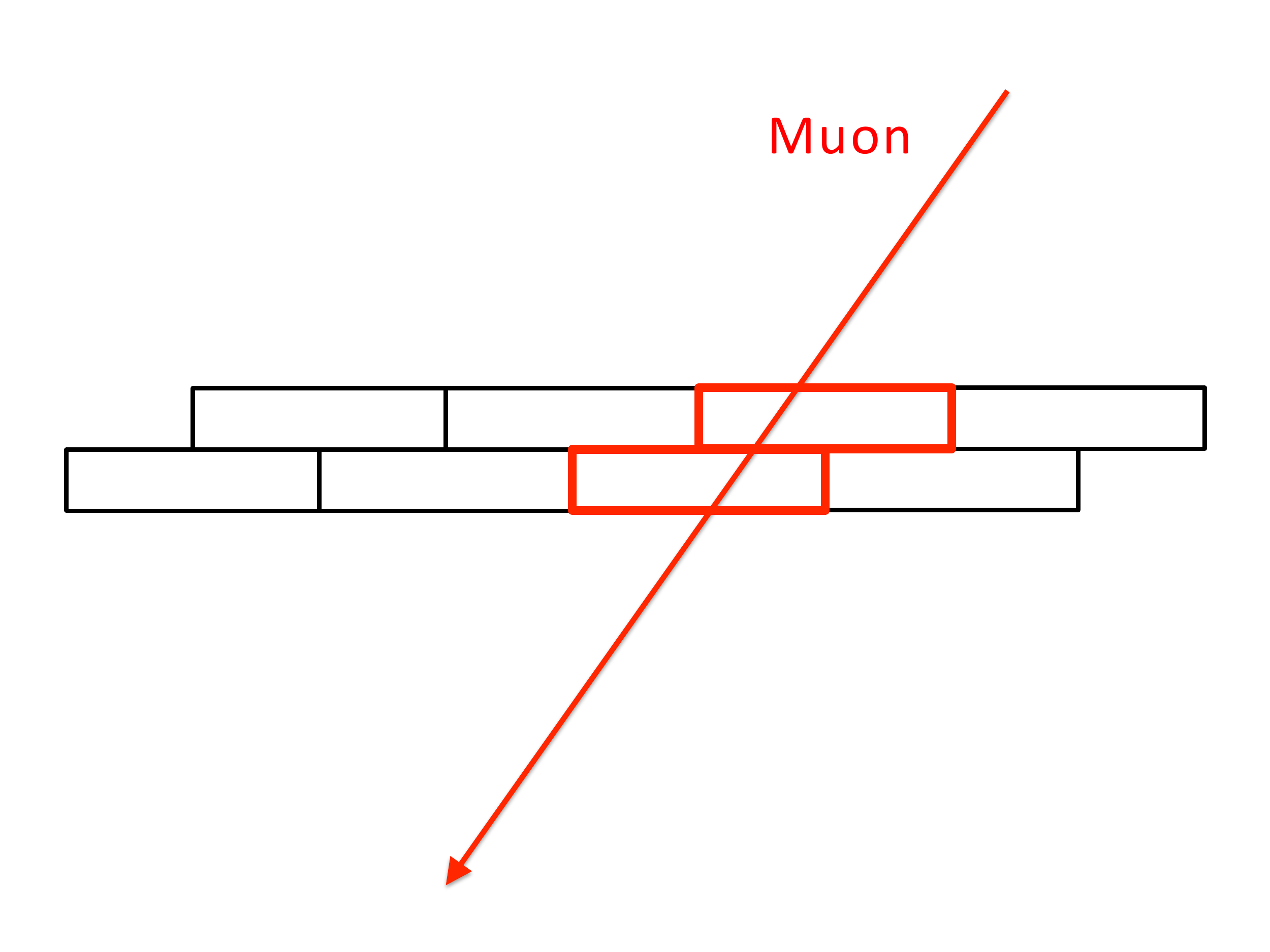} 
\caption{Example of a geometrically overlapping hit produced by a muon.}
\label{fig:bilayer}
\end{figure}

 The analysis technique and the muon event selection lean heavily on the so-called \emph{bilayer hits}. 
These are pairs of single hits on scintillator bars that overlap with each other. A simple illustration of such an example is shown in Figure~\ref{fig:bilayer}. A through-going muon will always produce geometrically overlapping hits in the absence of inefficiencies. Actually, this rather simple and intuitive feature was taken into account when designing and constructing this muon detector. In particular, the very specific orientation of scintillator bars in a bilayer was motivated by the decision to search for combinations of single hits in bars that geo\-metrically overlap. This choice helps to significantly reject background events from radioactivity or other instrumental noise (e.g. dark current, electronic and optical cross-talk). 
An event in the detector is classified as a muon when there are 8 geometrically overlapping hits; two for each of the 4 modules. In this way, most of the natural background radioactivity is discarded. A radiation $\gamma$ can fake a muon geometrically overlapping hit in our detector via double Compton scattering. A $\gamma$ can deposited energy similar to a MIP in two geometrically overlapping strips. This probability is around $\sim$5\% and it is greatly suppressed by the event selection requiring 8 geometrically overlapping strips.
This was verified through a dedicated Monte Carlo study and also by taking data with artificial radioactive sources. Radioactive sources were used, producing a range of $\gamma$s between 0.5 and 10~MeV. There were no $\gamma$ events with 8 geometrically overlapping hits. The same procedure was used to determine the background due to neutrons. Also in this case the requirement of 8 hits reduces to a negligible level the background due to neutrons. 
\par
\begin{figure}[h!]
\centering
\begin{tabular}{ c c }
\includegraphics[width=6.5cm, height=4.2cm]{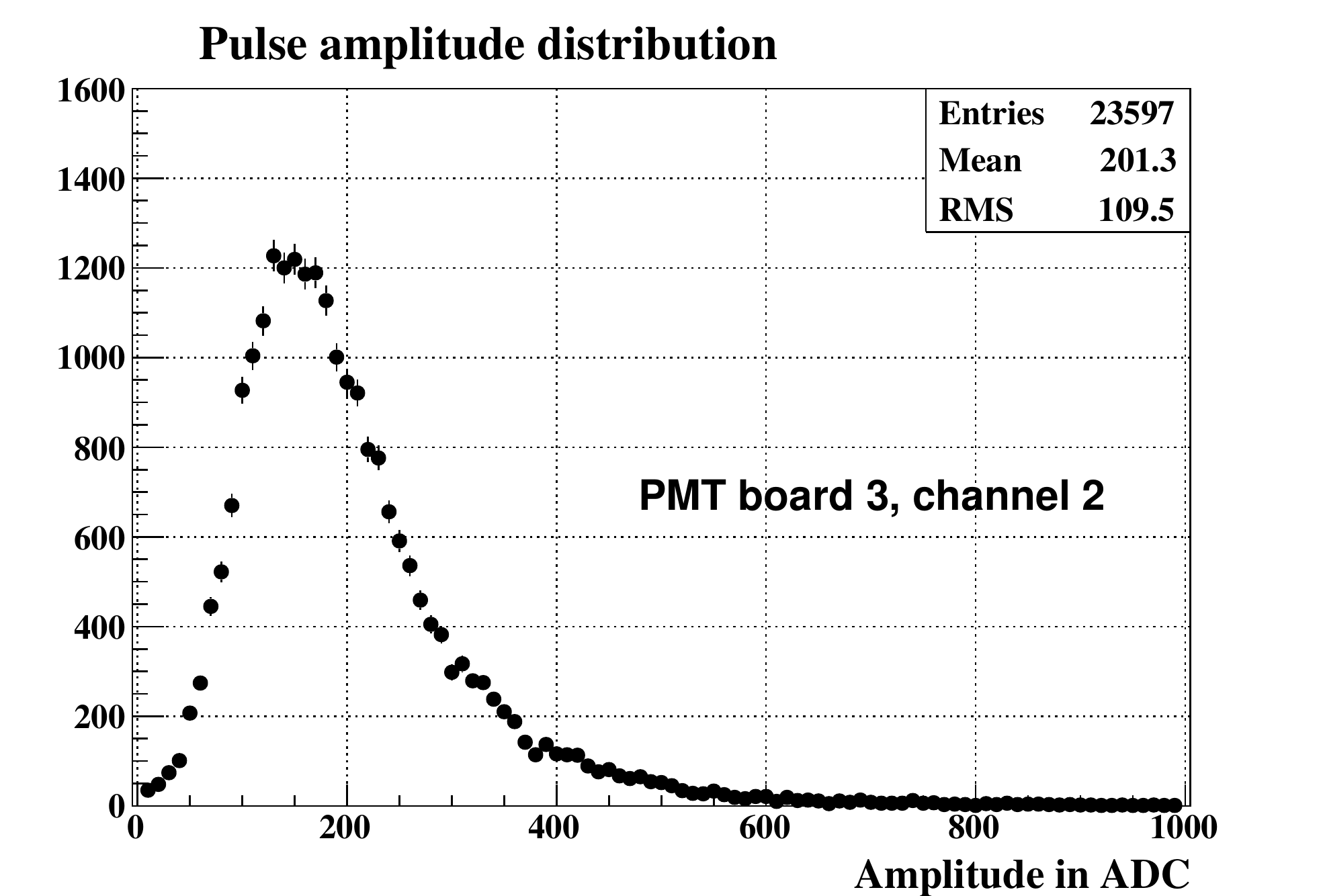} & \includegraphics[width=6.5cm, height=4.2cm]{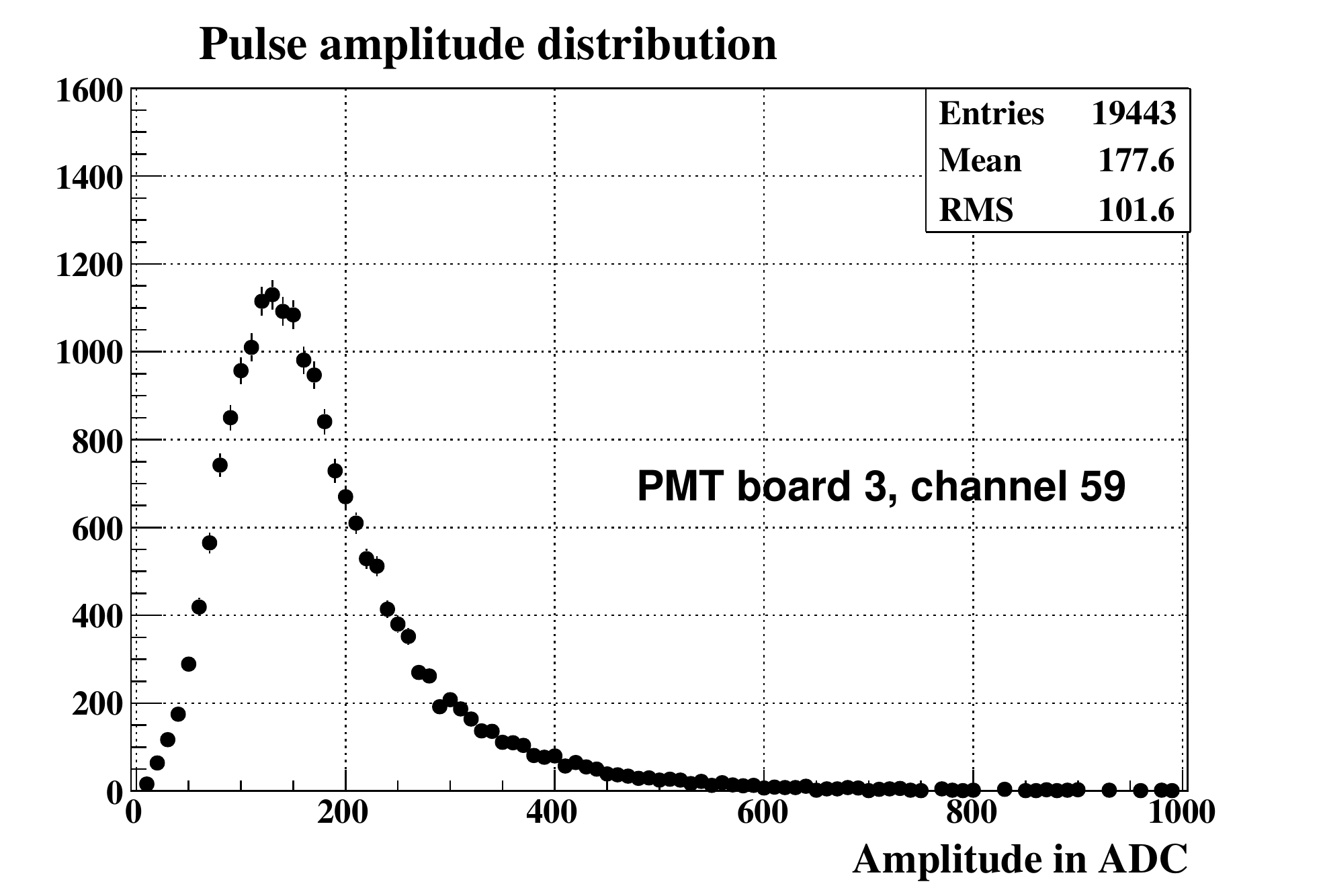} \\
\includegraphics[width=6.5cm, height=4.2cm]{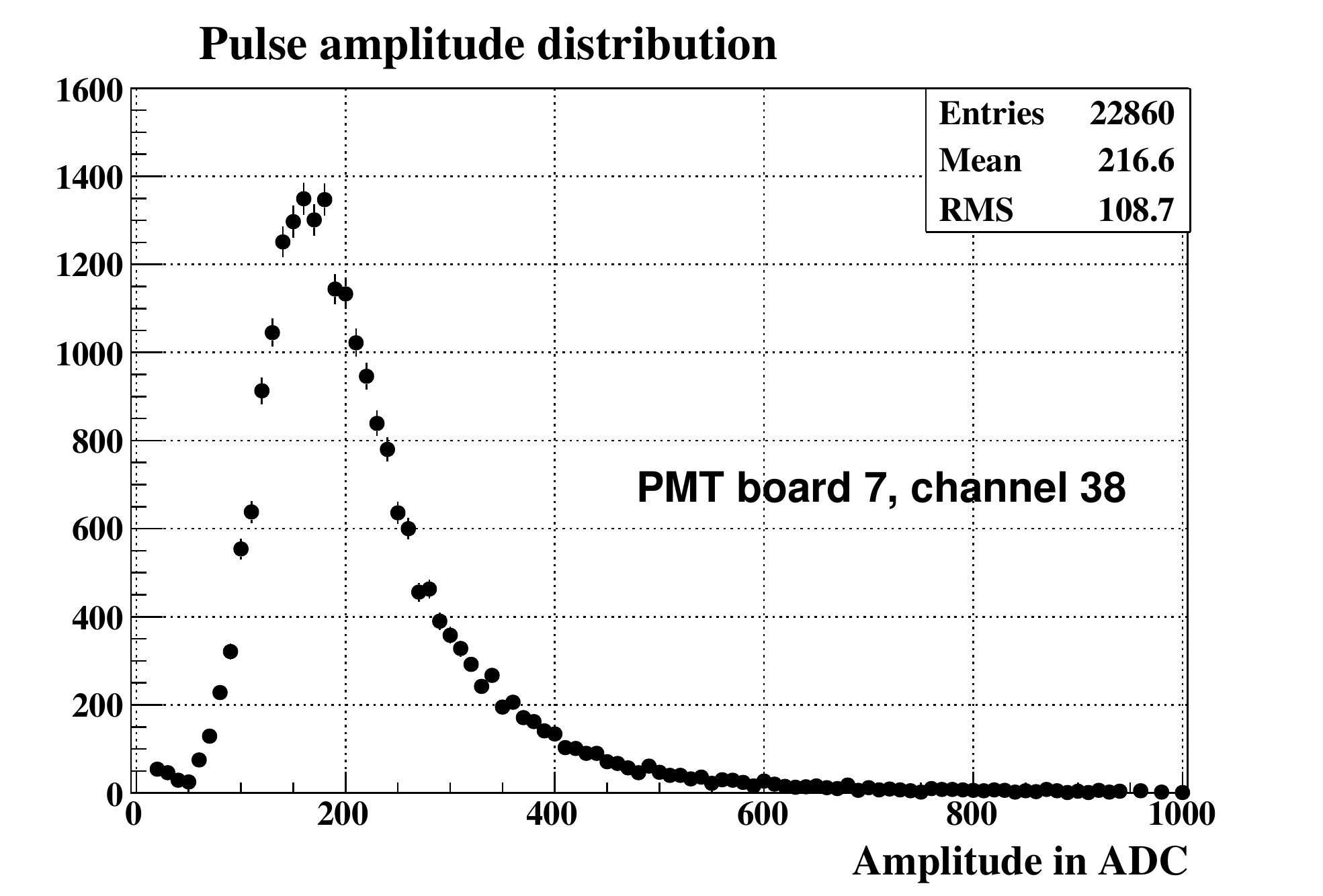} & \includegraphics[width=6.5cm, height=4.2cm]{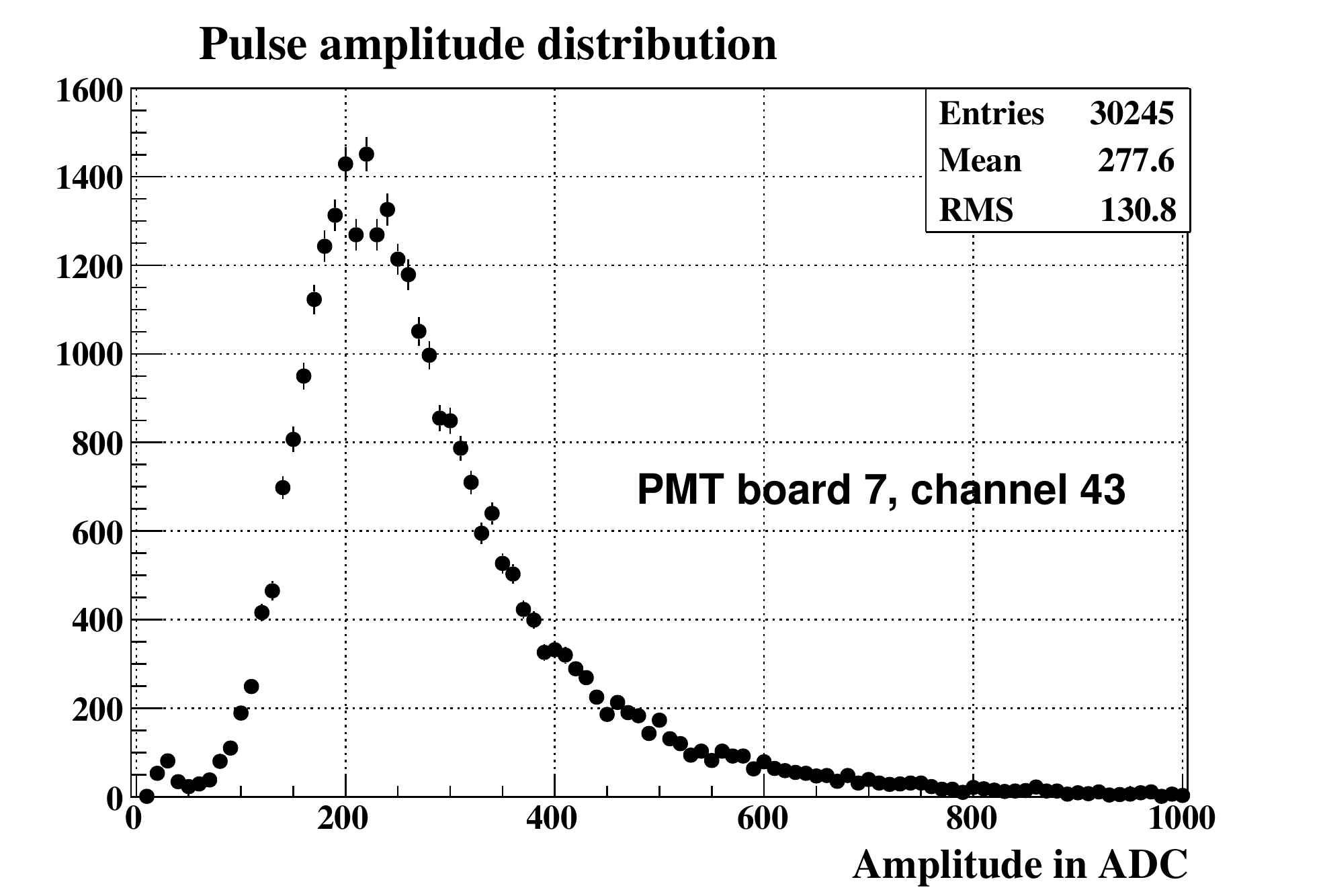}
\end{tabular}
\caption{Sample of pulse amplitude distributions in 4 of the 8 channels after applying the geometrical requirement.}
\label{fig:adc}
\end{figure}

 Figure~\ref{fig:adc} shows the pulse amplitude distributions from some random channels of the detector after the 8 hit requirement. 
For these plots, data from the runs taken at Virginia Tech were used. In all cases the pulse distributions have a \emph{Landau-like} shape as it is expected for scintillator response to a minimum ionizing particles. Note also that a ``soft'' exponential component originating from natural radioactivity has been cut off due to the strict selection cut. The curves of Fig.~\ref{fig:adc} correspond to muons that cross the entire detector volume. 
\subsection{Detector calibration}
\begin{figure}[t!]
\centering
\begin{tabular}{ c c }
\includegraphics[width=7.3cm, height=5.2cm]{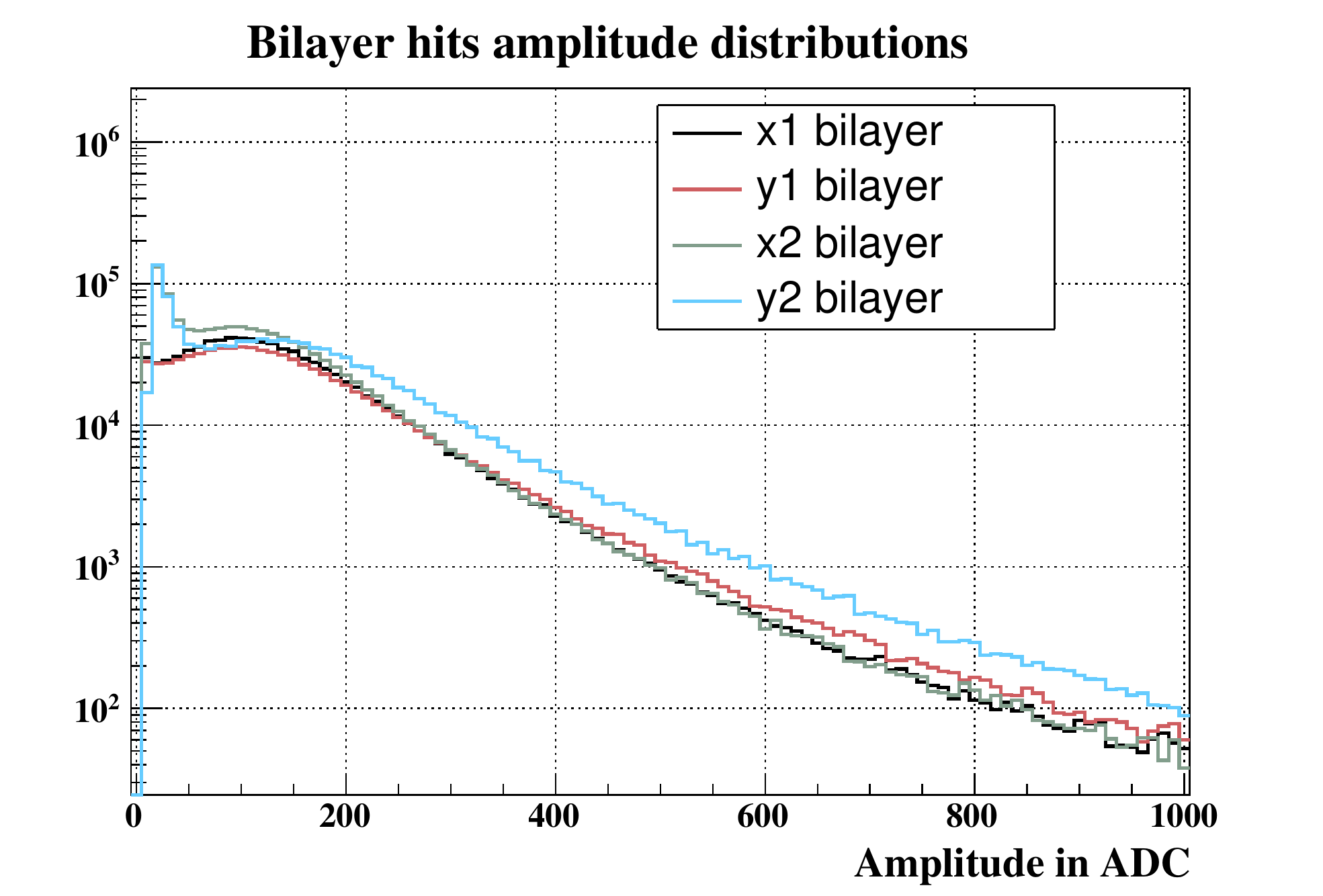} & \includegraphics[width=7.3cm, height=5.2cm]{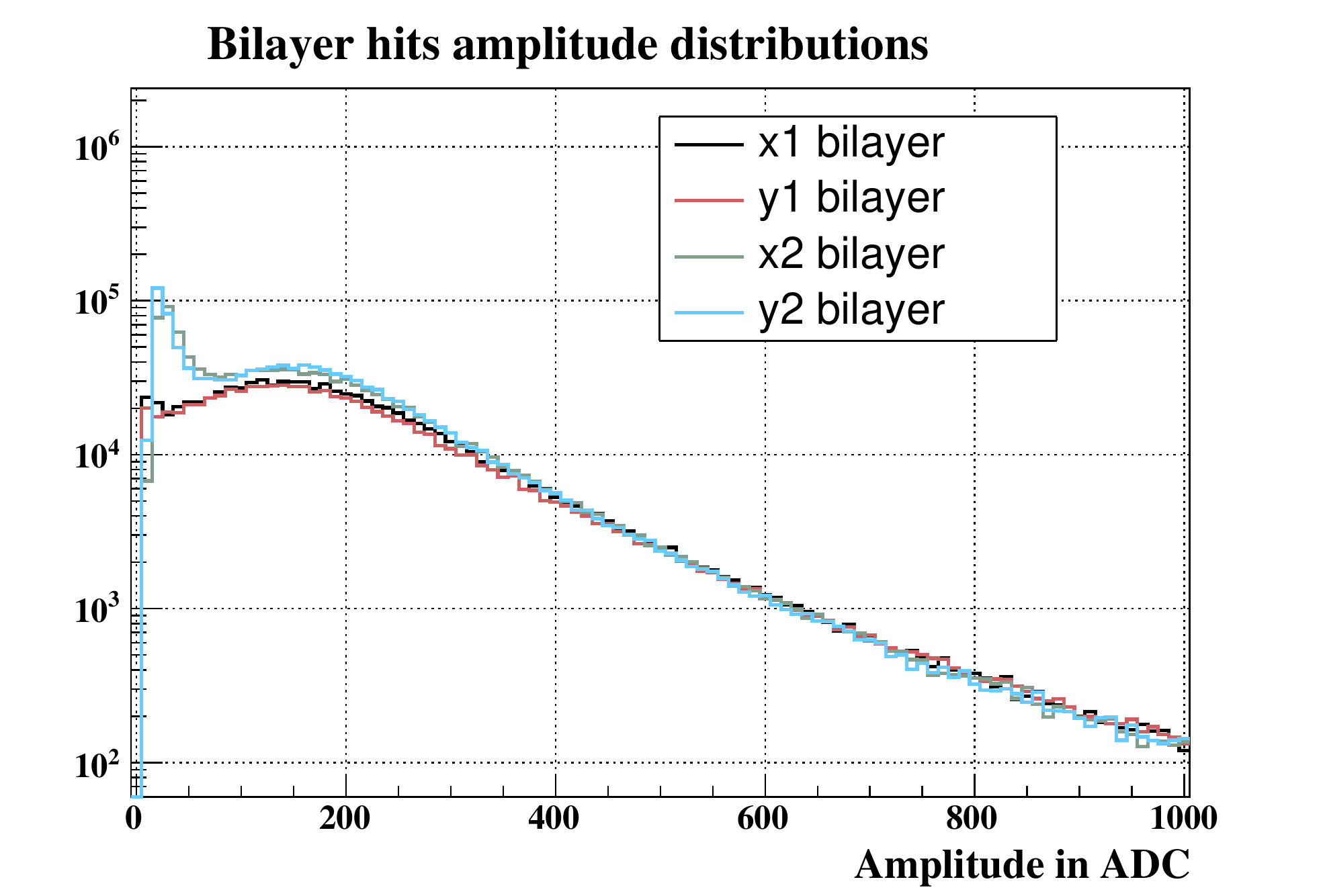} \\
\end{tabular}
\caption{Amplitude distributions in the $x_1$, $y_1$, $x_2$ and $y_2$ bilayer modules before (left) and after (right) calibration. For these plots the amplitude pairs from single bilayer hits were used. A threshold of 1/3 P.E above the pedestal value was used to generate the ADC distribution on the left and right panels.}
\label{fig:rad}
\end{figure}

%
The detector has been calibrated in two different ways. First, muon events were selected using just the 8 hits geometrical requirement. The ADC distributions for each of the MAPMT channels were plotted and then a correction factor was applied to equalize the means of all of those distributions. The decision taken, was  to arbitrarily equalize the mean of the ADC distributions to 200~ADC. Each channel was then rescaled in the software according to the calculated correction constant as in Eq.~\ref{eq:gain_constant}.
\begin{align}
\textrm{( channel-wise scaling factors )}_\alpha = \frac{200 \ \textrm{ADC} }{ \textrm{( MAPMT channel; mean ADC )}_\alpha }
\label{eq:gain_constant}
\end{align}
where $\alpha$ is a generic index referring to a specific channel in the detector and the denominator of Eq.~\ref{eq:gain_constant} is the mean ADC of the $\alpha$-th channel before this first calibration step. 
In this calibration step, the mean ADC values were used because, for low statistic samples (as was the case in KURF) the maximum of the ADC distribution was not easy to acquire in all channels, 
while the mean could be easily deduced from the statistics of the sample.
Of course, thirty-two scaling factors are obtained; one for each channel that reads out a scintillator strip. After that, all channels have muon signal distributions with the same mean. 
%
%
%
%
%
A second calibration is also applied. During the second calibration, the pulse amplitude distribution from all channels\footnote{This is the sum of the pulse amplitude distributions of all thirty-two channels.} is fitted with a gamma function. A gamma function is chosen instead of a Landau distribution 
for its simplicity. The ratio of the mean of the gamma distribution fit and 200~ADC is then used as a common rescaling factor to all channels. 
The second calibration has the effect of reducing the spread in the ADC distribution of the various channels. 
The effects of these two calibrations are showed in Fig.~\ref{fig:rad}.
After the calibration steps, the R.M.S. of the amplitude distribution from all channels narrows down by $\sim$25\%. Additionally, this recipe appears to be important in the extraction of the final rate results as shown in Ref.~\cite{ron}, removing any dependence of the muon rate on the software threshold utilized. 
%
%
\enlargethispage{-1\baselineskip}
\par This calibration technique appears to be particularly reliable when dealing with data samples that have limited statistics; these calibration steps were repeated for each location. 
The fluctuations in the High Voltage between data taking locations were determined to be about 10\%. To take this into account a site by site calibration was performed. 
Figure~\ref{fig:rad} shows the amplitude distributions (ADC) from those events in the four bilayers that have hits in overlapping bars before (left) and after (right) this data self-calibration. 
Those histograms were filled with the ADC values of those channels that geometrically overlap for those events with a bilayer hit in a bilayer. 
In each bilayer hit, two ADC values were filled  in the corresponding histogram (those of the two channels hit).
In the case of two bilayer hits four ADC values are filled in that histograms. 
In Figure~\ref{fig:rad}, $x_1$ and $y_1$ correspond to the two top bilayer modules of the detector and  $x_2$ and $y_2$ to the bottom ones. 
The same convention will used in the following parts of this article. 
One can easily observe the differences between these distributions of Figure~\ref{fig:rad}(left) and (right). 
After calibration, the spectra differ only in the region below $\sim$20~ADC ($\sim$1.2~P.E.) that corresponds to natural radioactivity. 
Beyond this limit the four curves very closely follow each other since the origin of this part is common; \emph{crossing muons}. 
Note that there is no pedestal in the plots of Figure~\ref{fig:rad} since a P.E. threshold of 1/3 P.E. has been applied. 
It should also be noted that bilayers $x_2$ and $y_2$ seem to have levels of natural radioactivity higher than those of $x_1$ and $y_1$. 
This could originate from the fact that those bilayers are at the bottom of the detector and accumulate more $\gamma$s from the ground. 
Differences in the performance of the two MAPMTs (e.g. dark current) or the bilayers $x_1$, $y_1$ and $x_2$, $y_2$ could also contribute to this effect. 
\par The muon rate is calculated requiring 8 hits geometrically overlapping and applying the two calibration factors as described above. 
The runtime for each of the runs is computed by looking at the difference between the first and last data packet recorded on disk. Our measured muon rates do not account for the inefficiencies due to High Voltage fluctuations and gain fluctuations. 
To account for these effects, the hit tagging efficiencies in each bilayer are separately estimated. Each time a muon crosses the detector there is a probability that no bilayer hits are registered in one of the four modules due to inefficiencies. 
The estimation of the muon rate can be corrected for that. The four module efficiencies are calculated through the following method. 
First  a particular module was chosen, for example the $y_1$, and only events with geometrically overlapping hits above the 4 P.E. threshold were selected in all the other modules. 
That is a very strict condition that only muons will satisfy. In a perfect detector there would be at least a bilayer hit above the minimal 1/3 P.E. threshold that would be recorded by the $y_1$ module as well. 
The percentage of cases in these 4 P.E. three bilayer events that indeed have also a bilayer hit at $y_1$ at a given P.E. threshold give us the hit tagging efficiency of $y_1$ at that threshold value. 
To be more precise, with the convention to represent as $N$$( \  x_1^{1/3} \  )$ the number of events with a bilayer hit at 1/3 P.E. (5.7 ADC) in $x_1$, the $y_1$ efficiencies are given by the formula:
\begin{align}
\epsilon_{y_1}^{1/3} = \frac{ N( \ x_1^{4} \wedge x_2^{4} \wedge y_2^{4} \wedge y_1^{1/3}  \ ) }{ N( \ x_1^{4} \wedge x_2^{4} \wedge y_2^{4} \ ) }    \label{eq:eff}
\end{align}
where $\wedge$ is the wedge logical $AND$ symbol and $ N( \ x_1^{4} \wedge x_2^{4} \wedge y_2^{4} \ )$ refers to the number of events with 4~P.E. (68 ADC) bilayer hits at $x_1$, $x_2$ and $y_2$. 
In a similar way $N( \ x_1^{4} \wedge x_2^{4} \wedge y_2^{4} \wedge y_1^{1/3}  \ )$ represents those triggered events with 4 P.E. bilayer hits at  $x_1$, $x_2$, $y_2$ and a 1/3 P.E. bilayer hit at $y_1$. $\epsilon_{x_1}^{1/3}$, $\epsilon_{x_2}^{1/3}$ and $\epsilon_{y_2}^{1/3}$ can be calculated through a simple substitution of the proper arguments in Equation~\ref{eq:eff} and the total efficiency of the detector at 1/3 P.E. threshold is given by the product:
\begin{align}
\epsilon^{1/3} = \epsilon_{x_1}^{1/3} \cdot \epsilon_{y_1}^{1/3}  \cdot \epsilon_{x_2}^{1/3}  \cdot \epsilon_{y_2}^{1/3}
\end{align}
The efficiency errors can be determined by properly propagating the statistical error in the number of events in Eq.~\ref{eq:eff} and it is reported together with the module efficiencies in Tab.~\ref{table:eff}.
\begin{table}[t!]
\centering
\begin{tabular}{| c | c |}
\hline
\hspace*{1cm} Bilayers   \hspace*{1cm} & \hspace*{1cm} Efficiencies ( \% ) \hspace*{1cm} \\ \hline
$x_1$  & 75.2 $\pm$ 0.1 \\
$y_1$  & 84.4 $\pm$ 0.1 \\
$x_2$  & 81.2 $\pm$ 0.1 \\
$y_2$  & 96.0 $\pm$ 0.1 \\
\hline
\end{tabular}
\caption{Bilayer efficiencies.}
\label{table:eff}
\end{table}
The total detector efficiency is $\sim$60\% requiring 8 geometrically overlapping strips at $1/3$ P.E. threshold. That efficiency is calculated integrating over all angles.
It should be remarked that the efficiencies in modules $x_1$ and $x_2$ are lower than those in the other two modules. That is due to the longer fiber used to connect these modules. The fiber bending radius is in this case also smaller and contribute to a decrease in the light collection efficiency. Another difference is in the photocathode efficiency of the MAPMT. 
A more efficient selection scheme can be implemented but in this analysis the choice made was to reduce the detector efficiency, increasing the purity of the muon sample. 
\par
\begin{figure}[t!]
\centering
\begin{tabular}{ c c }
\includegraphics[width=7.6cm, height=5.5cm]{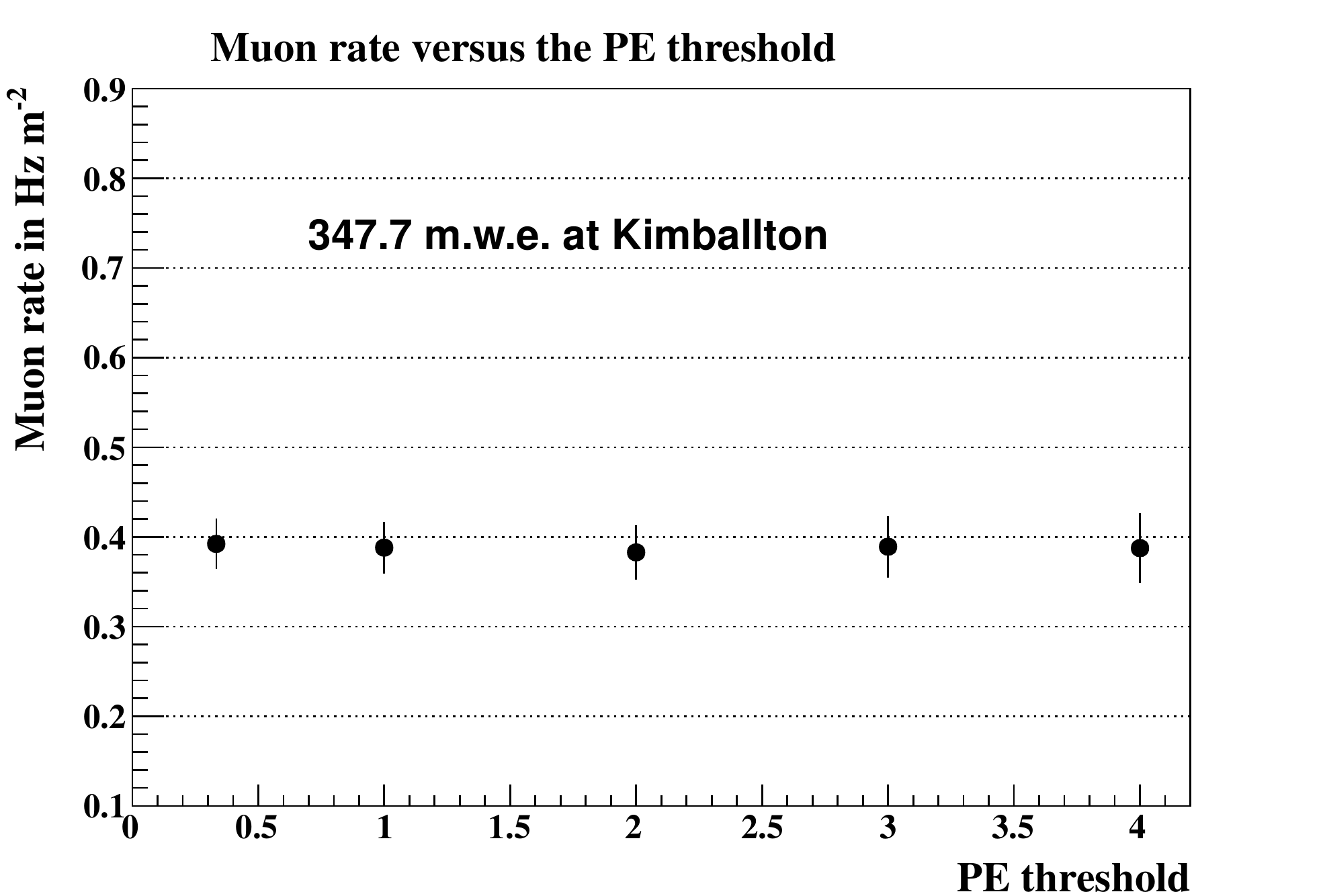} & \includegraphics[width=7.6cm, height=5.5cm]{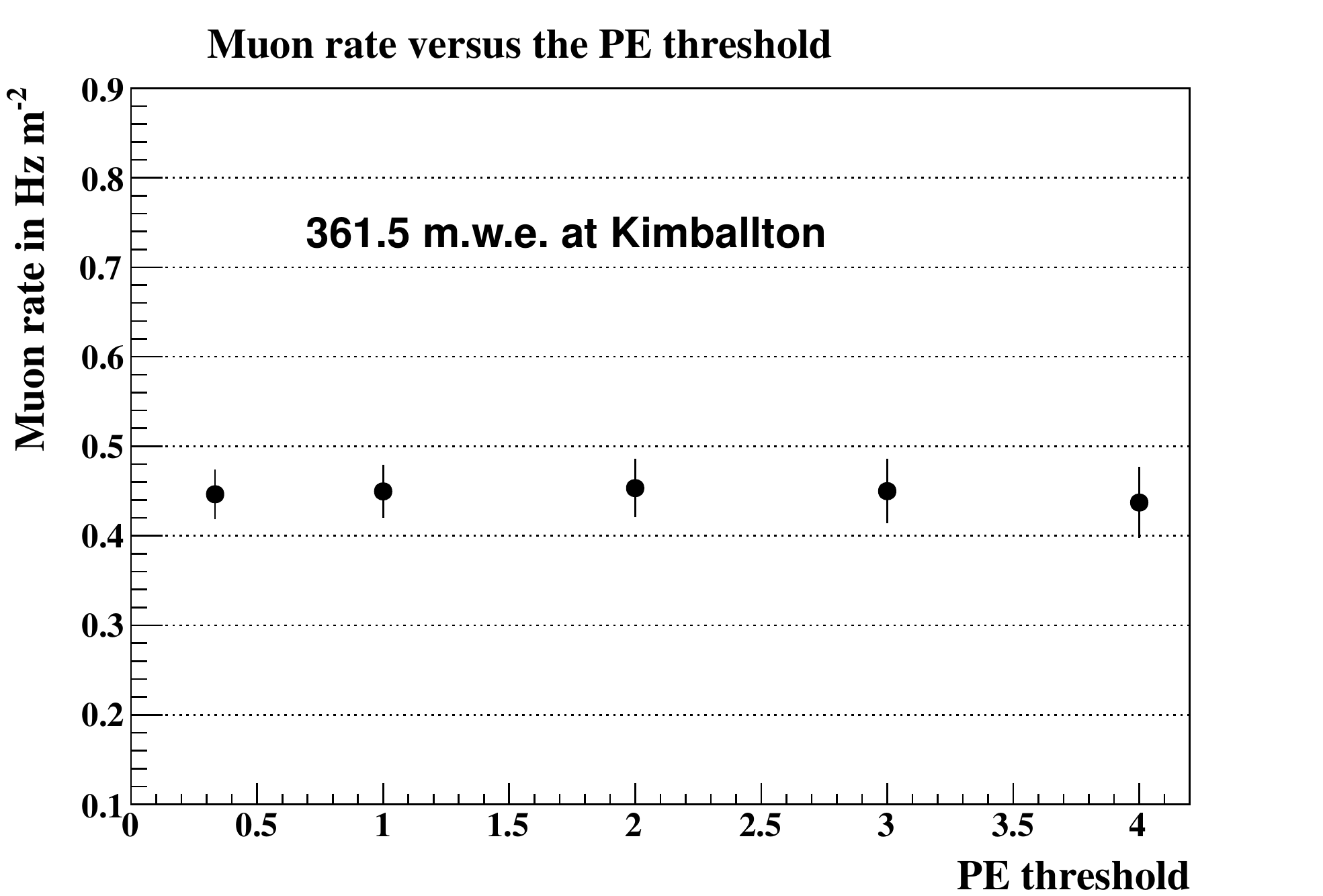} \\
\end{tabular}
\caption{Muon rate in the detector, as obtained from the four bilayer events and corrected for efficiencies, versus the P.E. threshold at two positions in KURF. The rate in both cases does not depend on the threshold applied.}
\label{fig:thres}
\end{figure}

 The same procedure applies to compute the efficiencies for the other thresholds. They were found them to be 56\%, 49\%, 41\% and 32\% respectively for 1, 2, 3 and 4 P.E. thresholds. As expected, the efficiency decreases as the threshold increases. The rate uncorrected for efficiency decreases as well as function of the P.E. threshold but the corrected muon rate as shown in Eq.~\ref{eq:muon_rate_corrected} remains constant.
\begin{align}
R_\mu^{x} = \frac{ N( \ x_1^{x} \wedge y_1^{x} \wedge x_2^{x} \wedge y_2^{x} \ ) } {\Delta T \cdot \epsilon^{x} }
\label{eq:muon_rate_corrected}
\end{align}
$\Delta T$ is the run length.
Figure~\ref{fig:thres} shows the extracted muon rate as a function of the P.E. threshold for two of the locations in the KURF. As can be seen in both plots the muon rates calculated after the efficiency corrections are constant within experimental errors. 
\par The final muon rate in Hz m$^{-2}$ is obtained scaling the measured rates for the effective horizontal area of the detector that is 0.175$\times$0.175~m$^2$. 
The effective detector area was determined considering the overlapping area of the four bilayers. 
Losses of events due to acceptance effects in the detector were taken account in the final systematics. 
In particular, the muon angular distributions at each site of the measurements were simulated and the amount of muons that are not selected 
due to the limited acceptance of the detector were found to be smaller than 10\% in all cases. 
For the final systematics the conservative value of 15\% was chosen to account as well the small increase in the effective area of the 
detector due to secondary particles ($\delta$ electrons, bremsstrahlung $\gamma$s) or instabilities in the electronics caused by the 
performance of the generator used to power up our setup inside the mine. 
Table~\ref{table:rate} summarizes the results for each of the data taking locations. 
These are the muon fluxes (Hz m$^{-2}$) at these locations, integrated over all zenith angles, through an horizontal surface and the threshold used was 2 P.E. 
This value was used because it appears to be the most appropriate to remove impurities in the selected samples. 
The errors quoted in Tab.~\ref{table:rate} come from the joint statistical and systematic uncertainties of the measurements.   
Having treated in a rather conservative way the acceptance of the detector for inclined muons, the numbers of Tab.~\ref{table:rate} correspond to the absolute muon rates in those locations within, of course, experimental errors. 
\begin{table}[t!]
\centering
\begin{tabular}{| c | c |}
\hline
\hspace*{1cm} Rock overburden ( m.w.e. )   \hspace*{1cm} & \hspace*{1cm} Muon rate ( Hz m$^{-2}$ ) \hspace*{1cm} \\ \hline
  0     &   170 $\pm$ 26 \\
   292 &   0.57 $\pm$ 0.09 \\
   307 &    0.59 $\pm$ 0.09 \\
   328 &    0.54 $\pm$ 0.09 \\
   340 &   0.42 $\pm$ 0.07 \\
   348 &    0.38 $\pm$ 0.07 \\
   358 &   0.35 $\pm$ 0.06 \\
   362 &  0.45 $\pm$ 0.08 \\
   377 &   0.36 $\pm$ 0.06 \\
 1,450 &  0.010 $\pm$ 0.002 \\
\hline
\end{tabular}
\caption{Muon flux results. }
\label{table:rate}
\end{table}
\section{Flux Prediction and Scaling with the Overburden}\label{sec:pred}
\enlargethispage{-1\baselineskip}
In parallel with the flux measurements,  the expected muon rate at various locations within KURF were evaluated, using well-established flux and muon propagation models. 
This work was done in two different ways. First, an analytical computation was performed to give us the muon flux assuming a specific rock overburden.
 For the simplicity of the calculation, two easily manageable hill profiles were chosen so that the numerical integration can be performed having an analytical formula for the overburden profile. 
These two models, termed as flat hill and perfect mountain, are expected to hold the essentials of this calculation since most of the experimental sites lie somewhere between these two extremes. Finally, a more detailed work was done using MUSIC \cite{music1, music2, music3}. MUSIC is a very robust muon propagation code that is known to model accurately many of the physical processes involving the interaction of muons and matter. For this latter work the exact topological profile of KURF was used. 
\par In all cases the muon directional intensity at sea level was taken from the Reyna parameterization~\cite{reyna}:
\begin{align}
 \frac{ d N_\mu}{dtdAd\Omega d p} = \cos^3\theta \cdot [ \ c_1\cdot \zeta^{-1( \ c_2 + c_3 \log_{10}\zeta+c_4\log^2_{10}\zeta + c_5\log^3_{10}\zeta\ ) } \ ] \label{eq:reyna}
\end{align}
where the left side of Eq.~\ref{eq:reyna} corresponds to the number of muons ($d N_\mu$) per second ($d t$) per square meter ($d A$) per steradian ($d \Omega$) between a specific momentum range ($ p\  - \  p + d p $).
$p$ is the muon momentum in GeV/c, $\theta$ the is muon zenith angle and $\zeta = p \cdot \cos\theta$. 
The term in the brackets corresponds to the muon vertical flux when $\theta=0^\textrm{o}$. The constants in Eq.~\ref{eq:reyna} were taken from fits to the world data and Reyna quotes the numbers:
\begin{align}
c_1= 0.00253, \quad c_2 = 0.2455, \quad c_3 = 1.288, \quad c_4 = -0.2555 \quad \textrm{and} \quad c_5 = 0.0209
\end{align}
These values were used for all momenta, but it should be understood that the Reyna model is only valid in the 1.0 -  2,000.0/$\cos\theta$ GeV/c region. 
For the analytical flux calculations, the minimum threshold energy values in each depth were taken from the formula:
\begin{align}
E_{min} = \epsilon ( \ e^{x^\prime/\xi} - 1 \ ). \label{eq:thresE}
\end{align}
$x^\prime$ is the slant depth in m.w.e., depending on the actual muon momentum direction, $\epsilon=500.0$ GeV and $\xi=2,500.0$~m.w.e. \cite{bible}. 
\par In the analytical models, the integration of the Reyna formula was done assuming a different energy threshold for each zenith angle direction. 
This threshold was calculated using Eq.~\ref{eq:thresE} and using two different overburden profiles to acquire the slant depth $x^\prime$. 
First, a totally flat overburden profile was assumed:
\begin{align}
x^\prime = x/\cos\theta
\end{align}
where $x$ is the vertical rock overburden in m.w.e. This will be referred as the \emph{flat hill} model. 
A second calculation was performed, where a real mountain was approximated with a perfect hemisphere and the detector is located at its center. 
In this case the slant depth was the same in all directions: 
\begin{align}
x^\prime = x, 
\end{align}
and the model was most appropriately called as the the \emph{perfect mountain}. 
Similar considerations have been  put forward also in \cite{landlike}. 
\begin{table}[t!]
\centering
\begin{tabular}{| c | c | c | c | c |}
\hline
\hspace*{0.0cm} Rock overburden  \hspace*{0.0cm} & \hspace*{0.0cm} Flat hill \hspace*{0.0cm} & \hspace*{0.0cm} Perfect mountain \hspace*{0.0cm} & \hspace*{0.0cm} MUSIC \hspace*{0.0cm} & \hspace*{0.0cm} Measurements \hspace*{0.0cm} \\ 
                                                ( m.w.e. )                              & ( Hz m$^{-2}$ )                                                & ( Hz m$^{-2}$ )                                                                   &  ( Hz m$^{-2}$ )                                      &  ( Hz m$^{-2}$ )       \\[0.5ex] 
\hline \
   292 &   0.56$^{+0.07}_{-0.06}$   & 1.11$^{+0.11}_{-0.10}$ & 0.65$^{+0.08}_{-0.06}$ & 0.57 $\pm$ 0.09 \\[0.5ex]
   307 &     0.51$^{+0.06}_{-0.05}$ & 1.01$^{+0.10}_{-0.09}$ & 0.54$^{+0.06}_{-0.06}$ & 0.59 $\pm$ 0.09 \\[0.5ex]
   328 &    0.44$^{+0.05}_{-0.04}$  & 0.88$^{+0.09}_{-0.08}$ & 0.50$^{+0.06}_{-0.05}$ & 0.54 $\pm$ 0.09 \\[0.5ex]
   340 &    0.41$^{+0.05}_{-0.04}$  & 0.82$^{+0.09}_{-0.08}$ & 0.43$^{+0.05}_{-0.04}$ & 0.42 $\pm$ 0.07 \\[0.5ex]
   348 &   0.39$^{+0.05}_{-0.04}$   & 0.79$^{+0.09}_{-0.08}$ & 0.40$^{+0.04}_{-0.04}$ & 0.38 $\pm$ 0.07 \\[0.5ex]
   358 &   0.36$^{+0.05}_{-0.04}$  & 0.74$^{+0.08}_{-0.07}$ & 0.35$^{+0.05}_{-0.03}$ &0.35 $\pm$ 0.06 \\[0.5ex]
   362 &   0.35$^{+0.04}_{-0.04}$  & 0.72$^{+0.08}_{-0.07}$ & 0.42$^{+0.05}_{-0.04}$ &0.45 $\pm$ 0.08 \\[0.5ex]
   377 &   0.32$^{+0.04}_{-0.03}$   & 0.66$^{+0.08}_{-0.07}$ & 0.33$^{+0.04}_{-0.03}$ & 0.36 $\pm$ 0.06 \\[0.5ex]
 1,450 &  0.0089$^{+0.0016}_{-0.0013}$ & 0.025$^{+0.004}_{-0.003}$ & 0.0067$^{+0.0012}_{-0.0010}$ & 0.010 $\pm$ 0.002 \\[0.5ex]
 \hline
\end{tabular}
\caption{Results from theoretical predictions side-by-side with the experimental measurements. }
\label{table:pred}
\end{table}
Table~\ref{table:pred} shows the results from both calculations, side-by-side with the experimentally measured muon fluxes. The results from MUSIC are also included in the same table. For all the predictions a 5\% uncertainty was taken on the rock overburden due to the uncertainty in the rock density and uncertainties in the survey of the mine as explained in Sec.~\ref{sec:det}. 
Our measurements should lie somewhere between the two analytical models that should be considered as lower and upper limits since most sites lie somewhere between these two extremes. 
On the other hand, the MUSIC simulations take advantage of the actual topological map of KURF and are expected to be more precise. As seen in the Tab.~\ref{table:pred}, the measurements in all cases are within the flat hill and perfect mountain assumptions, while the MUSIC predictions are in better agreement with the data. 
\par
\begin{figure}[!t]
\centering
\includegraphics[width=13.65cm, height=9.5cm]{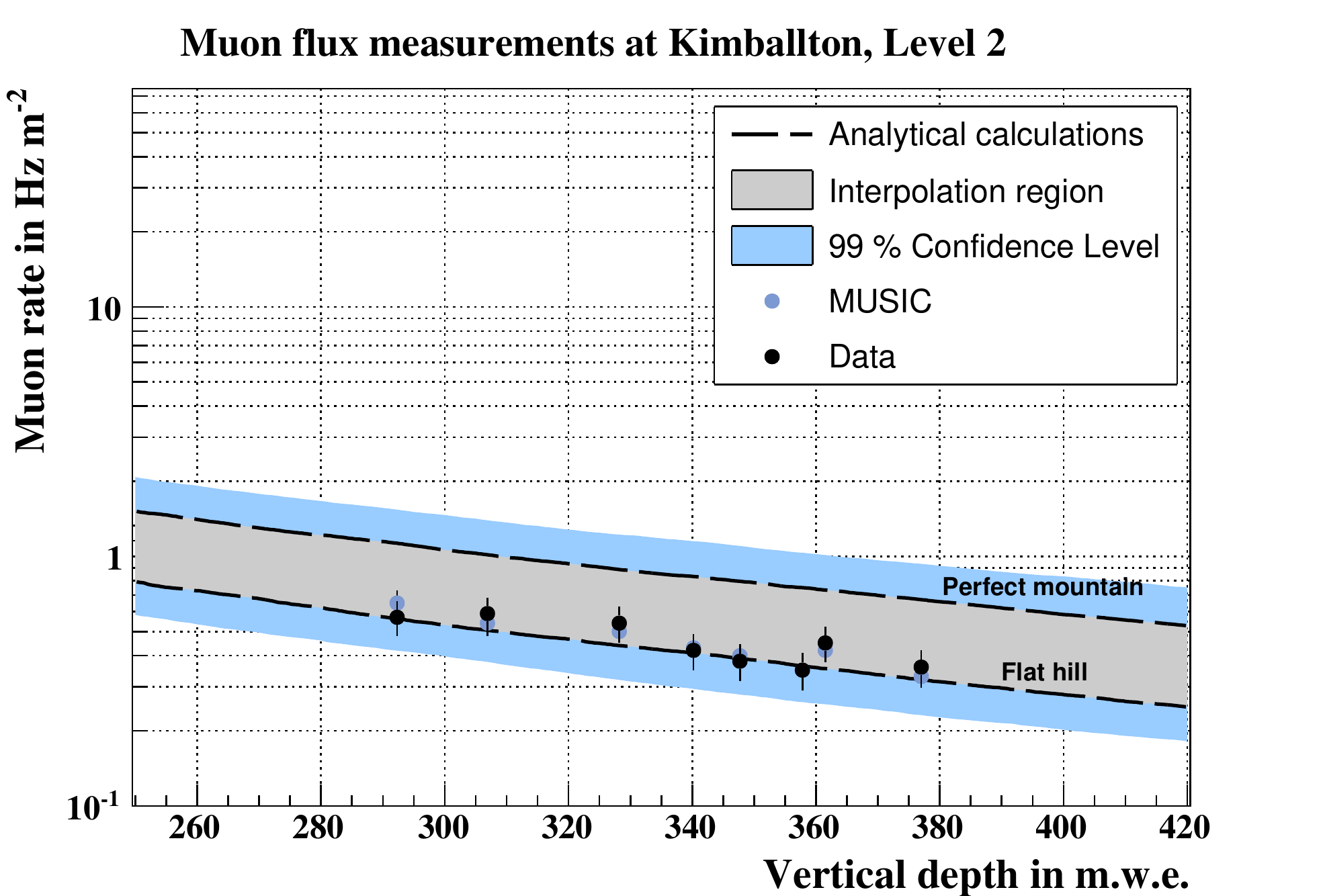}
\caption{Muon flux measurements and predictions at 2$^{nd}$ level in KURF.}
\label{fig:money}
\end{figure}

 In Figure~\ref{fig:money} the muon flux results are shown, overlapped with the various predictions for the 2$^{nd}$ level of KURF. 
The grey band denotes the interpolation region between the two analytical computations and in light blue the 99\% confidence belts.  
It should further be commented that a more significant difference between measurements and predictions is seen for the underground location at 1,450~m.w.e.
 This deviation arises from an incomplete knowledge of the exact overburden profile\footnote{Many underground cavities and tunnels exist in KURF that are not properly taken into account.}. 
To estimate the dependence of muon flux on the vertical overburden in KURF,  the data were fitted using the three parameter function: 
\begin{align}
I_{\mu}(\ x \ ) \ = \ A \left( \frac{x_0}{x} \right)^\alpha e^{-\frac{x}{x_0}}. \label{eq:macro}
\end{align}
$I_\mu(x)$ is muon flux (Hz m$^{-2}$) that crosses an horizontal surface at the underground vertical  of $x$ (m.w.e.) integrated over all angles and momenta. 
The same formula has been used many times in the past; for example in \cite{macro}. $\alpha$ is a pure number, $x_0$ is given in m.w.e. and $A$ in Hz m$^{-2}$.
A simple fit with $A$, $\alpha$ and $x_0$ unconstrained returns the parameter values :   
\begin{align}
A= 0.033 \pm 0.028\ \textrm{Hz m$^{-2}$}, \quad \alpha = 1.91 \pm 0.44, \quad \textrm{and} \quad x_0 = 1483 \pm 552 \ \textrm{m.w.e.} \label{eq:params}
\end{align}
The $\chi^2$ used for the fit was built as the following: 
\begin{align}
\chi^2 = \sum_{i=1}^8 \frac{( \ I_\mu^{(meas)}(x_i)-I_\mu^{(pred)}(x_i) \ )^2}{\sigma^2_{I_\mu^{(meas)}(x_i)}}, \label{eq:chi3}
\end{align}
where $I_\mu^{(meas)}(x_i)$ is the muon flux measured at the vertical overburden of $x_i$ inside KURF, 
$I_\mu^{(pred)}( x_i )$ is the muon flux predicted from Eq.~\ref{eq:macro} assuming some values for the parameters $A$, $\alpha$ and $x_0$, 
and $\sigma_{I_\mu^{(meas)}(x_i)}$ are the errors of $I_\mu^{(meas)}(x_i)$. 
The sum of Eq.~\ref{eq:chi3} extents over the eight measurements in Level 2 quoted in Table~\ref{table:pred}, $x_i$ = 292 - 377 m.w.e. and the best fit parameters were found minimizing the 
expression of Eq.~\ref{eq:chi3} with respect to $A$, $\alpha$ and $x_0$. 
It should be noted that the $\chi^2$ of Eq.~\ref{eq:chi3} has 5 degrees of freedom (NDOF) since we are fitting eight data points with a model of three free parameters, NDOF = 8-3 = 5. 

\begin{figure}[!t]
\centering
\includegraphics[width=13.65cm, height=9.5cm]{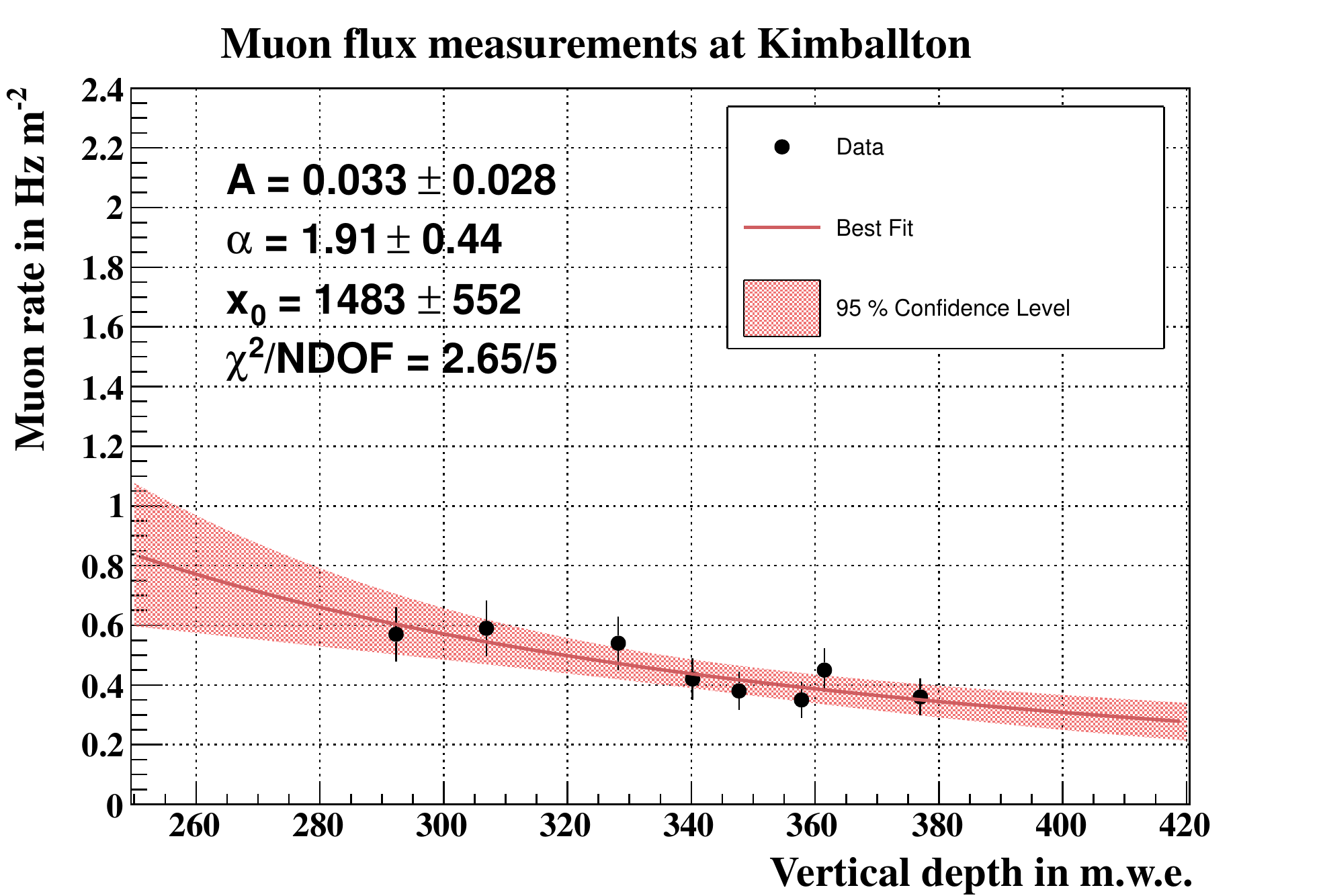}
\caption{Three parameter fit in the  2$^{nd}$ Level data. The best fit curve and the 95\% confidence intervals are also shown.}
\label{fig:fit}
\end{figure}

The relative high errors in the extracted numbers show that this measurement, with these errors, has limited sensitivity in the determination of  $A$, $\alpha$ and $x_0$. 
On the other hand the $\chi^2$ of the fit is not very bad, 2.65/5, and the final formula appears to parametrize well the scaling of the muon flux with at the overburden in Level 2 of KURF. Figure~\ref{fig:fit} shows the best fit curve and the 95\% confidence intervals. 
The above parameters were obtained from fits in the 2$^{nd}$ Level data only, but the validity of the final formula should be considered as more general. In particular Eq.~\ref{eq:macro} and \ref{eq:params} should be considered 
as a quick recipe to provide a good approximation of the muon rate in KURF given a specific vertical rock overburden. 
 Eq.~\ref{eq:macro} can be used to predict the muon flux at the 1,450 m.w.e. laboratory, and the result $
I_{\mu}(\ 1,450\ \textrm{ m.w.e. } \ ) = 0.0130 \textrm{ Hz m}^{-2}$,  is in excellent agreement with the measured value as presented in Sec.~\ref{sec:flux}. 
\section{Conclusions\label{sec:outro}}
We have measured the absolute muon flux at various levels in the Kimballton Underground Research Facility (KURF). The measurements were obtained using a small prototype muon detector made of plastic scintillator and readout by MAPMT and WLS fibers similar to the Double Chooz muon veto detector. Despite the limited area of our detector, we were able to take statistically significant measurements at the level 2 of KURF ($\sim$300 m.w.e.) and compared them with the predictions of well-established and tested muon rate prediction models like MUSIC. The MUSIC prediction depends on the vertical overburden data that was used as input to the simulation. We found that the measurements were in very good agreement with the flux predictions of the MUSIC simulation package. We have then determined an empirical formula to determine the flux at various overburdens and extrapolated our predictions to much deeper location in the mine, 1,450 m.w.e. and found that in excellent agreement with the measurements.
\acknowledgments
The authors would like to thank the Neutrino group of Columbia University and Nevis Laboratories for the construction and the assembly of the portable muon detector. 
We would also thank Lhoist for giving us access to the Kimballton mine and KURF laboratory. 
Special thanks go to Prof. B. Vogelaar, Dr. D. Rountree, Z. Lewis, T. Wright for helping us with the data taking in the mine. 
Additionally, we are grateful to Dr. V. Kudryavtsev and Prof. A. Tonazzo for many useful discussions concerning the muon flux predictions. 
Finally, we would like to thank Dr. S. Manecki for valuable comments and criticism on earlier versions of this article. 
\end{document}